\documentclass[9pt,conference]{IEEEtran}
\IEEEoverridecommandlockouts
\usepackage{cite}
\usepackage{amsmath,amssymb,amsfonts}
\usepackage{algorithmic}
\usepackage{graphicx}
\usepackage{textcomp}
\usepackage{xcolor}
\usepackage{url}
\usepackage{booktabs}
\def\BibTeX{{\rm B\kern-.05em{\sc i\kern-.025em b}\kern-.08em
    T\kern-.1667em\lower.7ex\hbox{E}\kern-.125emX}}

\makeatletter
\def\ps@IEEEtitlepagestyle{%
  \def\@oddfoot{\mycopyrightnotice}%
  \def\@oddhead{\hbox{}\@IEEEheaderstyle\leftmark\hfil\thepage}\relax
  \def\@evenhead{\@IEEEheaderstyle\thepage\hfil\leftmark\hbox{}}\relax
  \def\@evenfoot{}%
}
\def\mycopyrightnotice{%
  \begin{minipage}{\textwidth}
  \footnotesize
  \copyright~2024 IEEE. Personal use of this material is permitted.  Permission from IEEE must be obtained for all other uses, in any current or future media, including reprinting/republishing this material for advertising or promotional purposes, creating new collective works, for resale or redistribution to servers or lists, or reuse of any copyrighted component of this work in other works.
  \end{minipage}
}
\makeatother
    
\begin{document}

\title{A Non-Intrusive Neural Quality Assessment Model for Surface Electromyography Signals\\}

\author{\IEEEauthorblockN{Cho-Yuan Lee\IEEEauthorrefmark{1}\IEEEauthorrefmark{5},
Kuan-Chen Wang\IEEEauthorrefmark{2}\IEEEauthorrefmark{5}, Kai-Chun Liu\IEEEauthorrefmark{3}, Yu-Te Wang\IEEEauthorrefmark{5}, Xugang Lu\IEEEauthorrefmark{4}, Ping-Cheng Yeh\IEEEauthorrefmark{2}, and
Yu Tsao\IEEEauthorrefmark{5}}
\IEEEauthorblockA{
\IEEEauthorrefmark{1}School of Medicine, National Yang Ming Chiao Tung University, Taipei, Taiwan\\
\IEEEauthorrefmark{2}Graduate Institute of Communication Engineering, National Taiwan University, Taipei, Taiwan \\
\IEEEauthorrefmark{3}College of Information and
Computer Sciences, University of Massachusetts, Amherst, MA, 01003, USA\\
\IEEEauthorrefmark{4}National Institute of Information and Communications Technology, Tokyo, Japan \\
\IEEEauthorrefmark{5}Research Center for Information Technology Innovation, Academia Sinica, Taipei, Taiwan \\
Email: rick.109101013.md09@nycu.edu.tw,\ 
d12942016@ntu.edu.tw,\
kaichunliu@umass.edu,\\
yutewang@citi.sinica.edu.tw,\
xugang.lu@nict.go.jp,\
pcyeh@ntu.edu.tw,\
yu.tsao@citi.sinica.edu.tw
}
}


\maketitle

\begin{abstract}
In practical scenarios involving the measurement of surface electromyography (sEMG) in muscles, particularly those areas near the heart, one of the primary sources of contamination is the presence of electrocardiogram (ECG) signals. To assess the quality of real-world sEMG data more effectively, this study proposes QASE-net, a new non-intrusive model that predicts the SNR of sEMG signals. QASE-net integrates a one-dimensional Convolutional Neural Network (CNN) with a Bidirectional Long Short-Term Memory (BLSTM) layer and attention mechanisms, following an end-to-end training strategy. Our experimental framework utilizes real-world sEMG and ECG data from two open-access databases, the Non-Invasive Adaptive Prosthetics Database and the MIT-BIH Normal Sinus Rhythm Database, respectively. The experimental results demonstrate the superiority of QASE-net over the baseline method, exhibiting significantly reduced prediction errors and notably higher linear correlations with the ground truth. These findings show the potential of QASE-net to substantially enhance the reliability and precision of sEMG quality assessment in practical applications. 
\end{abstract}


\section{Introduction}
Surface electromyography (sEMG) is a technique popularly used to measure the electrical currents generated by muscle cell depolarization during contractions noninvasively. Numerous applications adopt sEMG to extract information of muscle activities, including the diagnosis of neuromuscular disorders~\cite{wimalaratna2002quantitative, hogrel2005clinical}, rehabilitation~\cite{campanini2020surface}, speech aids~\cite{vojtech2021surface}, prosthetic devices~\cite{farina2014extraction}, and human-computer interaction~\cite{zheng2022surface}. In these applications, sEMG recordings may be contaminated by electrocardiogram (ECG) when the measured muscles are near the heart~\cite{fraser2012removal, mak2010automated}. This contamination can compromise the performance of sEMG applications~\cite{zhou2007real}. Thus, developing accurate and automated sEMG quality assessment approaches is critical to addressing the ECG contamination issue.

To quantify the signal quality of real-world sEMG data lacking clean references, certain studies have resorted to expert visual assessments of sEMG quality ~\cite{mak2010automated,farago2022review}. However, manual assessment is expensive, time-consuming, and subjective in real-world implementation. Other studies have employed metrics that require a clean version of the signal as ground truth for accurately assessing sEMG distortion~\cite{ma2020emg,petersen2020removing}. These challenges show a requirement for an objective and automated sEMG quality assessment. Additionally, precise SNR estimation can improve the robustness of sEMG denoising algorithms. Some studies have discovered that estimating SNR can help improve denoising techniques by balancing noise reduction with potential signal distortion, enabling appropriate noise cancellation~\cite{fu2016snr,ge2022percepnet+}.

In recent years, neural networks (NNs) have been widely applied in sEMG-related research owing to their powerful modeling capability. These sEMG applications include noise removal~\cite{kale2009intelligent, wang2023ecg}, contaminant type identification~\cite{machado2021deep}, gesture recognition~\cite{wei2019multi, cote2019deep}, and disease classification~\cite{elamvazuthi2015electromyography, samanta2020neuromuscular}. A study has also used NNs for SNR estimation in sEMG contaminated by ECG~\cite{oo2020signal}. In this study, SNR estimation was based on a multilayer perceptron (MLP) with various hand-crafted features. The results of this paper showed that, among the various sEMG features, the waveform length (WL) extracted from raw sEMG waveforms emerged as the most suitable feature for SNR estimation under ECG contamination.  While this work has shown satisfactory results for SNR estimation using NNs, it should be noted that their method was validated solely on simulated sEMG data.

Different NN architectures with unique properties are designed to improve the representation of information in data, including convolutional neural networks (CNNs)~\cite{lecun1995convolutional}, recurrent neural networks (RNNs)~\cite{hochreiter1997long}, and attention mechanism~\cite{vaswani2017attention}. These NN models have advantages over MLPs in various signal processing tasks, particularly for time-series data analyses~\cite{fu2016snr,chiang2019noise}. This study proposed a more robust and accurate SNR estimation methodology for sEMG, termed QASE-net, which leverages a CNN-BLSTM model integrated with the attention mechanisms.  Instead of using hand-crafted features, QASE-net employs convolutional layers to extract features from raw waveform directly, along with an end-to-end training strategy. The results show that the proposed approach achieves a notable improvement in SNR prediction accuracy compared to the conventional method, especially at higher SNR levels. Crucially, our experimental data emulates real-world scenarios by encompassing genuine both sEMG and ECG data to validate the applicability of the proposed method.

\section{Related Work}
\subsection{SNR estimation of sEMG with ECG interference}
In ~\cite{oo2020signal}, Oo et al. developed SNR estimation methods for sEMG signals contaminated by ECG interference. Their approach involved the use of an MLP with hand-crafted features. The MLP architecture included two hidden layers comprising 20 neurons with a sigmoid activation function, followed by a linear output layer. Simulated sEMG data was corrupted by simulated ECG interference to train the SNR estimation model, while real ECG data was employed during testing. The SNR levels of the noisy sEMG signals in both the training and testing sets ranged from 0 to -20 dB, and the contaminated sEMG segments were normalized through standardization and energy-based normalization. The study investigated six hand-crafted features: WL, skewness, kurtosis, mean absolute value, zero crossing, and mean frequency. These features were extracted from the raw sEMG waveform and the fourth- and fifth-level outputs of discrete stationary wavelet transform applied to sEMG data. Experimental results demonstrated that the WL features from the raw waveform yielded the most precise SNR estimation, with a linear correlation coefficient (LCC) of 0.9663 between the actual and predicted SNR levels, where the WL feature is calculated as follows:
\begin{equation}
    \label{eq:WL}
    \text{WL} = \sum_{i = 1}^{N-1}|x[i+1] - x[i]|,
\end{equation}
where $N$ and $x[i]$ are the total length and the $i$-th sample of a segment of noisy sEMG signal, respectively. 

\subsection{NN-based signal quality assessment}
NN-based signal quality assessment has proven to be effective in various domains, including biomedical signal processing~\cite{zhang2019cascaded, kido2019novel, liu2023fgsqa} and speech analysis~\cite{fu2018quality, zezario2020stoi, zezario2022deep, kirton2023towards}. These techniques offer a non-intrusive (i.e., without needing a clean reference) means of evaluating signal quality, rendering them applicable to real-world scenarios where clean references are often unavailable. Moreover, these NN-based quality assessment approaches can be integrated with downstream systems to optimize the overall performance jointly. In addition to their high applicability, NN-based approaches can perform automated signal quality estimation, potentially yielding substantial cost savings compared to human-based subjective assessments.

Within the realm of NN-based signal quality assessment, various NN models, such as CNNs, RNNs, and the attention mechanism, are frequently incorporated into the model architectures. CNNs excel in extracting features from time-series data~\cite{zhang2019cascaded, kido2019novel, liu2023fgsqa}, leveraging convolutional filters to discern local relationships and capture hierarchical features. RNNs contain self-loops to retain information from previous time steps, making them adept at learning dependencies within sequential data. When combined with CNNs in signal quality assessment, RNNs can address the challenge of capturing long-term dependencies within the data~\cite{fu2018quality}. Among RNNs, Bidirectional Long Short-Term Memory (BLSTM) networks are frequently employed because they can mitigate the vanishing gradient problem and capture information from past and future time steps. Recent studies have demonstrated the prowess of combining CNN-BLSTM models with attention mechanisms in signal quality assessment~\cite{zezario2020stoi, zezario2022deep, kirton2023towards}. This design enables the model to autonomously extract features, capture temporal dependencies, and weigh the significance of feature maps for the final prediction. As a result, this architecture holds the potential to establish a more potent and data-driven approach in contrast to conventional hand-crafted features.

\section{Materials and methods}
\subsection{Public database of sEMG and ECG}
The open-access Non-Invasive Adaptive Prosthetics (NINAPro) database is used in this work~\cite{atzori2014electromyography}. This database recorded 12 channels of sEMG signals, acquired using Delsys Trigno electrodes placed on the upper limbs. The sampling rate of sEMG was 2 kHz. The NINAPro database comprised several subsets of data. In our experiment, we adopted the DB2 with sEMG recordings from 40 intact subjects, as it had the most subjects. Within DB2, we focused on two specific sessions, denoted as Exercise 1 and Exercise 2. These two sessions include 17 finger movements and 22 grasping and functional movements. Each movement was repeated six times for five seconds, followed by a three-second rest interval. Prior studies have utilized sEMG data from this database, employing filter preprocessing to obtain clean sEMG data~\cite{wang2023ecg,machado2021deep}.
 
For ECG interference, a PhysioNet database, MIT-BIH Normal Sinus Rhythm Database (NSRD), was adopted in this study~\cite{goldberger2000physiobank}. In MIT-BIH NSRD, 2-channel ECG recordings were collected from 18 healthy subjects, with a sampling rate of 128 Hz. In previous studies, the ECG signals from this database were used as the interference in sEMG for noise removal ~\cite{wang2023ecg} and noise-type identification~\cite{machado2021deep}.

\subsection{Data preparation}
As mentioned in the previous section, proper filtering is necessary to eliminate potential noise from the sEMG data within the NINAPro database. To achieve this, we employed a 4th-order Butterworth bandpass filter, with a passband ranging from 20 to 500 Hz, to process the sEMG signals.~\cite{machado2021deep,wang2023ecg}. Subsequently, the sEMG data were downsampled to 1 kHz and normalized by division with their maximum absolute values. Each sEMG signal was then partitioned into 10-second segments. Similarly, we processed the ECG recordings from the MIT-BIH NSRD database. These ECG signals were subjected to a 3rd-order Butterworth high-pass filter and a low-pass filter with respective cutoff frequencies of 10 Hz and 200 Hz.~\cite{xu2020comparative}.

For the training dataset, we adopted the sEMG segments in Channels 1 to 4, Exercise 1, from 24 subjects. For each sEMG segment, we randomly selected ECG signals from 12 subjects in the MIT-BIH NSRD as ECG interference. These ECG signals were then superimposed onto the clean sEMG segments as additive noise as done in previous studies~\cite{fraser2012removal, wang2023ecg}, at SNRs ranging from -15 to 0 dB, with an increment of 1 dB. The SNR is calculated as: 
\begin{equation}
    \label{eq:SNR}
    \text{SNR (dB)} = 10 \log_{10} \left( \frac{P_\text{signal}}{P_\text{noise}} \right),
\end{equation}
where $P_\text{signal}$ and $P_\text{noise}$ denote the power of the sEMG and ECG signals, respectively.

To evaluate the generalization capability of the proposed QASE-net, mismatch conditions were introduced between the training and testing datasets. For the validation set, we selected sEMG segments in Channels 5 and 6, Exercise 1, from subjects 25 to 32. The sEMG segments in Channels 9 and 10, Exercise 2, from subjects 33 to 40, were selected as our testing data. ECG from the remaining six subjects (18184, 19088, 19090, 19093, 19140, and 19830) were utilized as interference. These ECG signals were added to the clean sEMG segments at SNRs ranging from -15 to 0 dB, with an increment of 0.5 dB. In total, there were 137792, 43710, and 43648 segments of contaminated sEMG data prepared for training, validation, and testing, respectively.

\subsection{Network architecture of the proposed method}
\begin{figure}[bt!]
    \centering
    \includegraphics[height=\columnwidth]{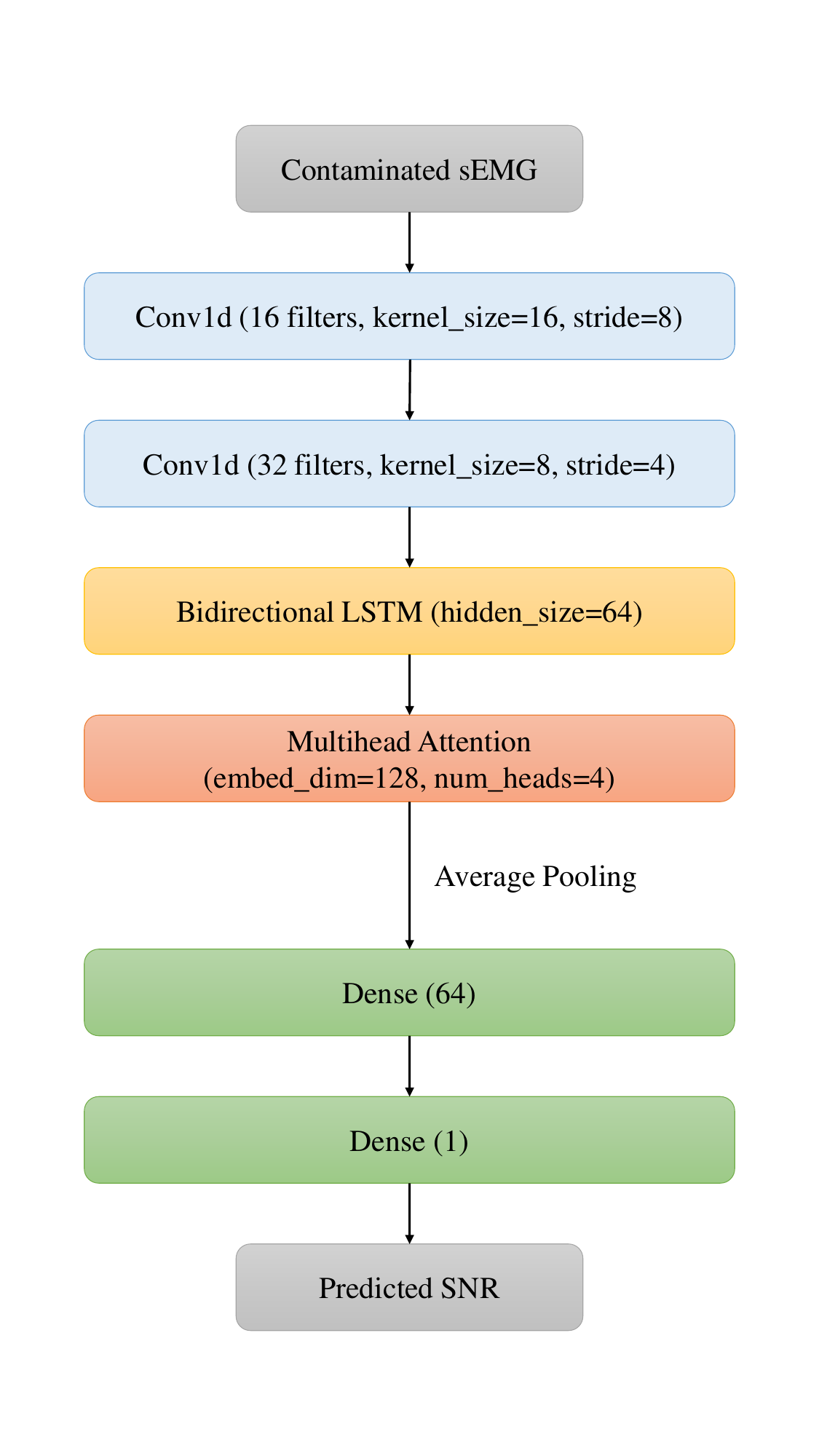}
    \caption{Architecture of the proposed method.}
    \label{fig:model}
\end{figure}

As discussed in Section 2, CNN, RNN, and attention mechanisms have been widely employed for signal quality assessment~\cite{zezario2020stoi, zezario2022deep, kirton2023towards}. This study adopted a similar architecture, and the model directly used raw sEMG waveform as its input. The structure of the proposed method is illustrated in Fig.~\ref{fig:model}, and the model parameters are estimated in an end-to-end training strategy. QASE-net begins with two one-dimensional convolutional layers to extract low-level features from the input sEMG waveform. The first convolutional layer consists of 16 filters with a kernel size of 16 and a stride of 8, and the second consists of 32 filters with a kernel size of 8 and a stride of 4. Each layer is followed by a batch normalization layer and a ReLU activation function. The feature maps produced by the CNN layers are then input to a BLSTM layer with 64 hidden units. Subsequently, the BLSTM output is processed by a multi-head self-attention layer with 4 heads and 128 embedding dimensions. Following the attention layer, the extracted features are aggregated through temporal average pooling, yielding a fixed-size representation. Lastly, the features are processed using two fully connected layers. The dimension of the hidden representation is reduced to 64, followed by a ReLU activation function. The final linear layer predicts the SNR value of the input sEMG signal.  

\subsection{Evaluation Metrics}
In this study, four evaluation metrics were used to assess the performance of QASE-net: the mean absolute error (MAE), mean squared error (MSE), linear correlation coefficient (LCC), and Spearman's rank correlation coefficient (SRCC). MAE and MSE measure the error between predictions and ground truth, with MSE assigning more penalties to more significant errors than MAE. These metrics are defined as follows:
\begin{equation}
    \label{eq:MAE}
    \text{MAE} = \frac{1}{m}\sum_{i=1}^{m} |s_e(i) - s_t(i)|,
\end{equation}
\begin{equation}
    \label{eq:MSE}
    \text{MSE} = \frac{1}{m}\sum_{i=1}^{m} (s_e(i) - s_t(i))^2.
\end{equation}
where $s_e(i)$ and $s_t(i)$ denote the estimated and true SNR, respectively. Lower values of MSE and MAE suggest better prediction performance. In contrast, LCC and SRCC evaluate the correlation between predictions and ground truth values. LCC is calculated as:
\begin{equation}
    \label{eq:LCC}
    \text{LCC} = \frac{\sum_{i=1}^{m}(s_t(i)-\bar{s_t})(s_e(i)-\bar{s_e})}{\sqrt{\sum_{i=1}^{m}(s_t(i)-\bar{s_t})^2}\sqrt{\sum_{i=1}^{m}(s_e(i)-\bar{s_e})^2}},
\end{equation}
SRCC measures the correlation between the ranked values of the two variables, making it less sensitive to outliers and indicating the strength of the monotonic relationship between actual and estimated SNRs. Higher values of LCC and SRCC suggest more accurate SNR prediction performance.

For baseline systems, we first followed ~\cite{oo2020signal} to build an SNR prediction model based on MLP with the WL feature, termed WL+MLP, in the following discussion. Since the frame size for extracting WL was not specified in \cite{oo2020signal}, we set the frame size to 200 ms, which yielded the best performance on our validation set. We further implemented two additional models, CNN and CNN-BLSTM, for comparison with our proposed QASE-net. The CNN model comprised five convolutional layers, while the CNN-BLSTM model consisted of three convolutional layers followed by two BLSTM layers. Both models incorporated temporal average pooling and two fully connected layers for SNR estimation.

\subsection{Implementation details}
We used the L2 loss as our cost function to train the neural assessment models. Given an sEMG segment, the loss was calculated as the mean square error between the predicted output and the ground-truth SNR values. The model parameters were optimized using the Adam optimizer. The batch size was set to 32. To achieve the best results on the validation set for each model, the WL+MLP was trained for 30 epochs with a learning rate of 0.001, while other models were trained for 20 epochs with a learning rate of 0.0001.\footnote{The code is available at {https://github.com/zylee-md/QASE-net}.} When provided with a noisy sEMG signal in the testing phase, the assessment network predicted its corresponding SNR value without requiring a clean reference, making it a non-intrusive model.

\section{Results and discussion}
\begin{figure}[th!]
    \centering
    \includegraphics[width=\columnwidth]{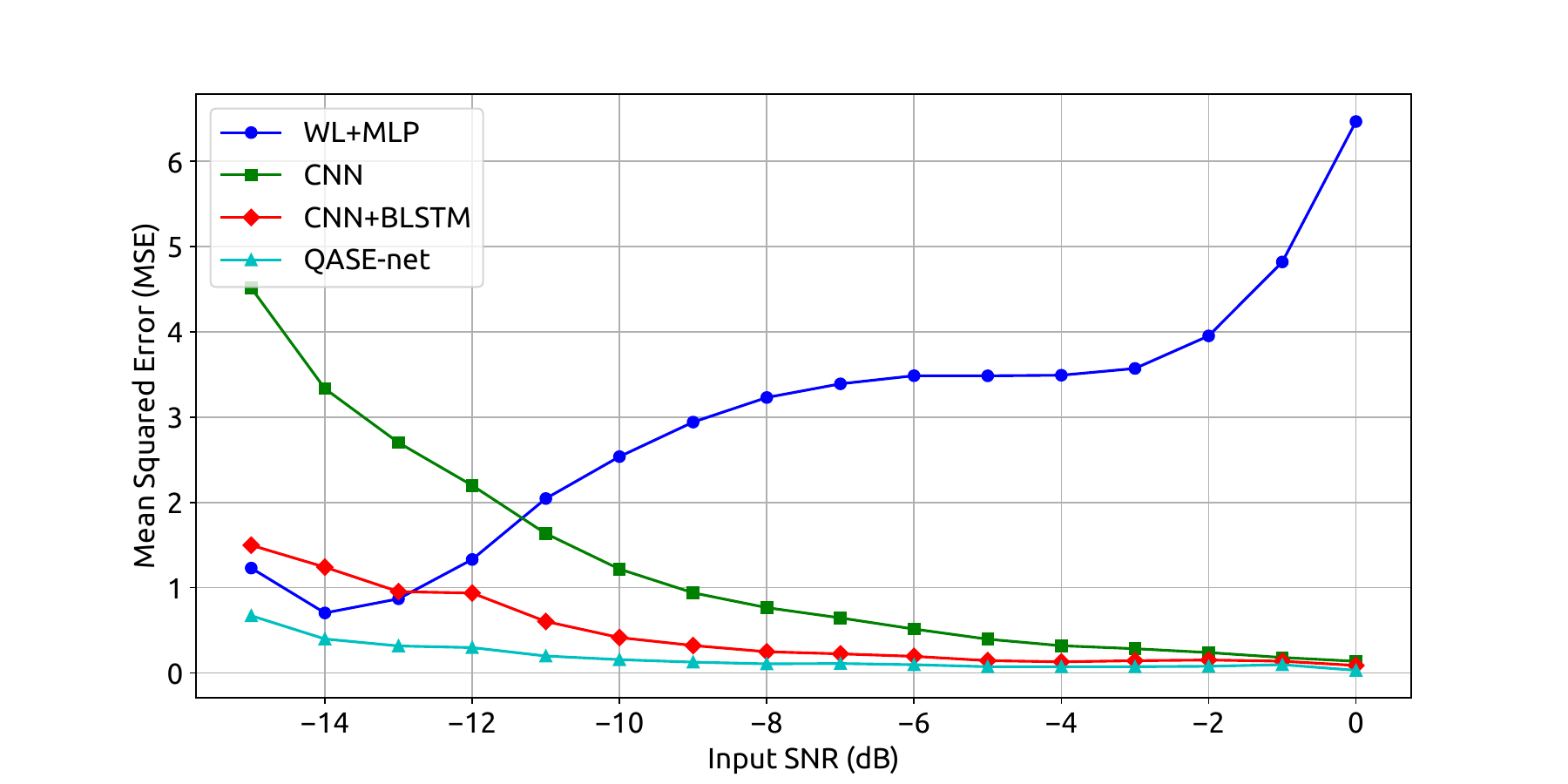}
    \caption{The MSE values of SNR predictions of QASE-net and comparative models under different SNR inputs.}
    \label{fig:mse}
\end{figure}

\begin{figure}[t!]
    \centering
    \includegraphics[width=\columnwidth]{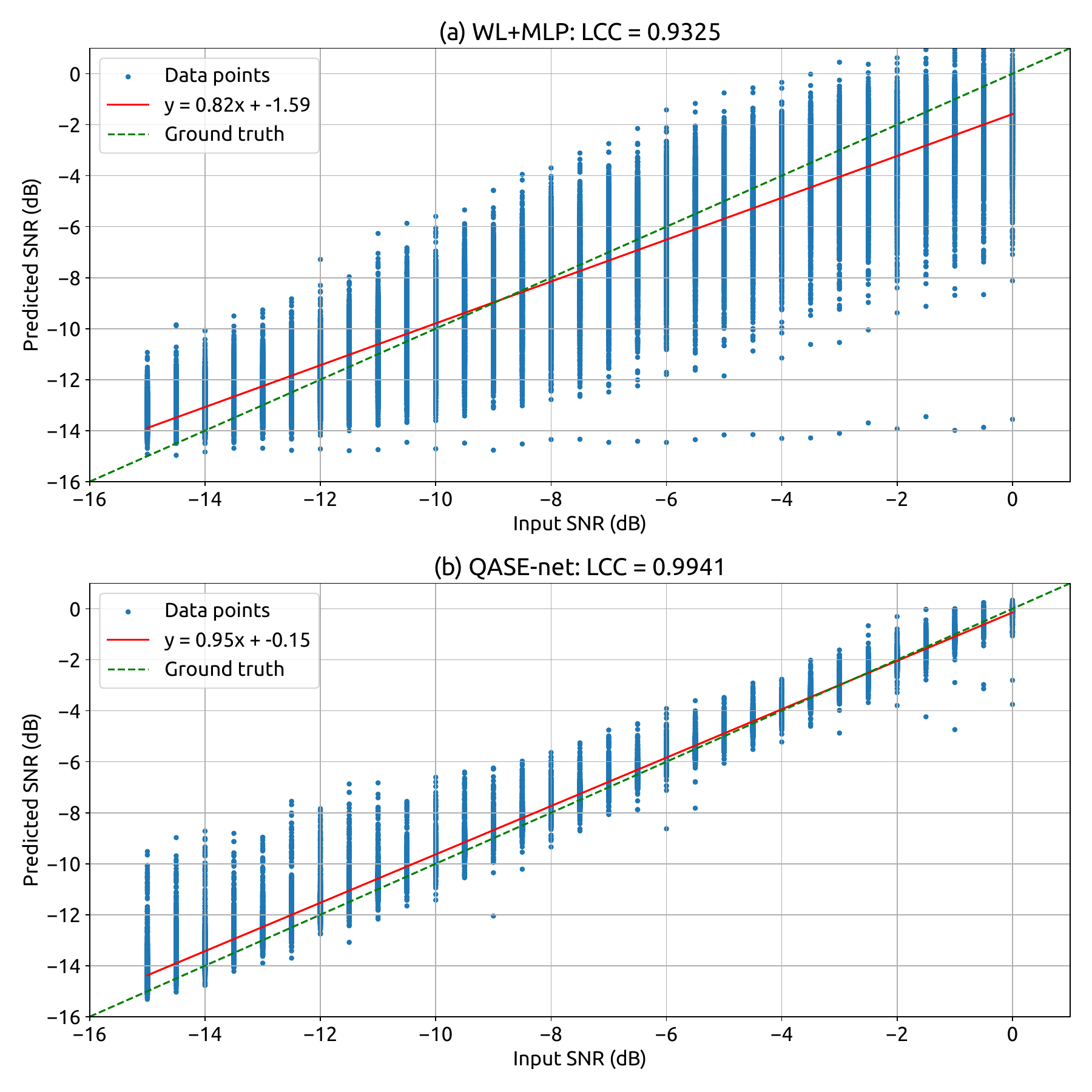}
    \caption{Scatter plots of the SNR predictions by (a) WL+MLP~\cite{oo2020signal} and (b) the proposed QASE-net.}
    \label{fig:scatter}
\end{figure}

\begin{figure}[t!]
    \centering
    \includegraphics[width=\columnwidth]{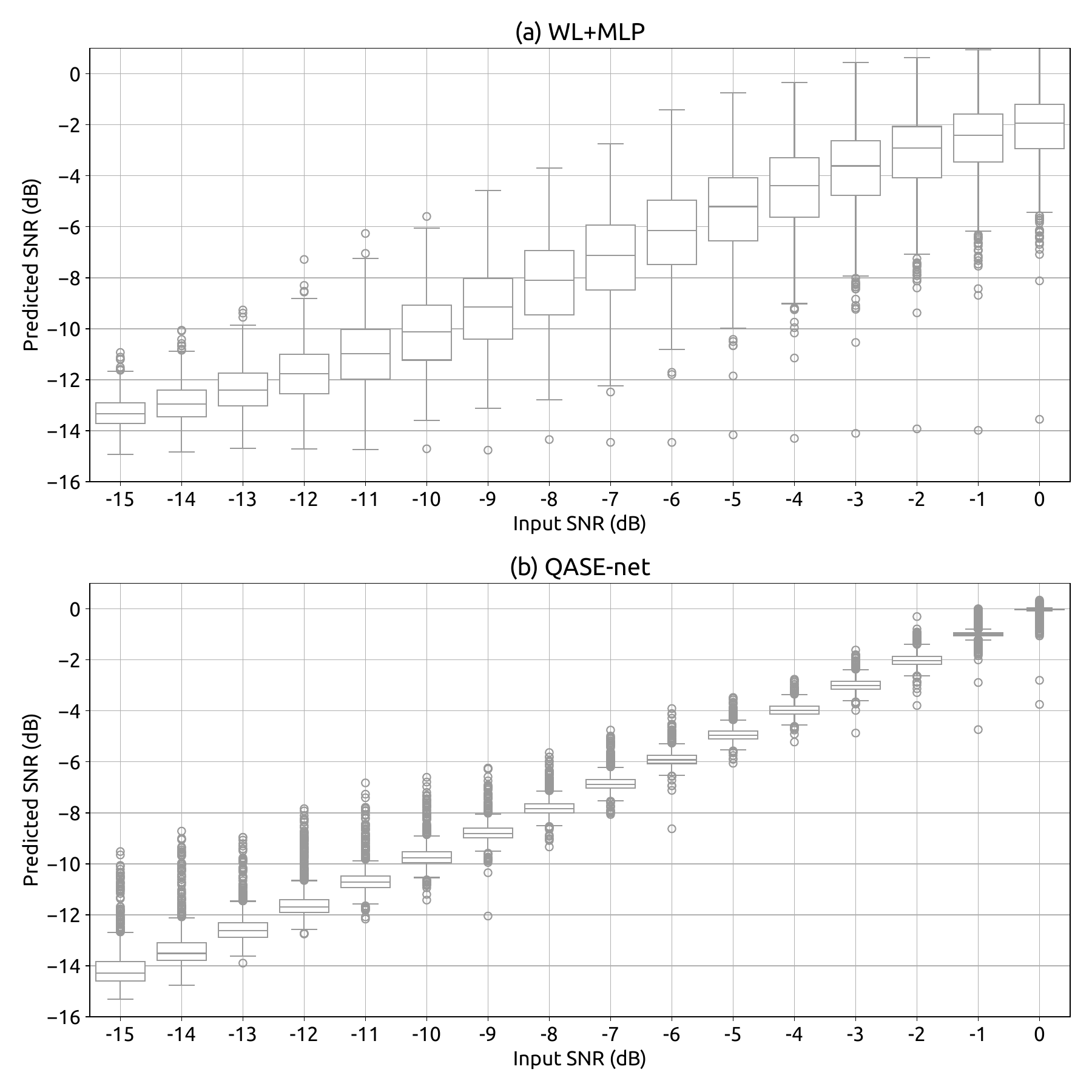}
    \caption{Box plots of SNR predictions by (a) WL+MLP~\cite{oo2020signal} and (b) the proposed QASE-net.}
    \label{fig:box}
\end{figure}

To ensure a thorough assessment of the model's performance, we conducted experiments 30 times with distinct seeds and calculated the standard deviation (SD) for each reported metric in Table~\ref{table:metrics}. We selected a particular seed and visualized its results, including the MSE values shown in Fig.~\ref{fig:mse}, the scatter plots showcased in Fig.\ref{fig:scatter}, and the corresponding box plots featured in Fig.\ref{fig:box}.

Table~\ref{table:metrics} shows the performance of QASE-net and other methods for comparison. The proposed QASE-net achieves the best performance with the lowest MAE and MSE and the highest correlation coefficients. Among the comparative methods, WL+MLP yields a reasonable correlation with the ground truth (LCC = 0.9352) but with relatively high MAE (1.3023) and MSE (2.7579). Next, QASE-net, CNN, and CNN+BLSTM consistently perform better than WL+MLP in all metrics, indicating that using convolutional filters for automated feature extraction leads to more accurate estimations than hand-crafted features. Moreover, CNN+BLSTM outperforms CNN, demonstrating the advantage of combining convolutional layers with BLSTM to capture temporal dependencies more effectively.

\begin{table*}[th!]
    \caption{The overall performance of SNR prediction methods.}
    \label{table:metrics}
    \centering 
    \begin{tabular}{ccccc}
     \bf{Method} & \bf{MAE} $\downarrow$ & \bf{MSE} $\downarrow$ & \bf{LCC} $\uparrow$ & \bf{SRCC} $\uparrow$\\
    \toprule
    WL+MLP ~\cite{oo2020signal}     & 1.3023 $\pm$ 0.0157 & 2.7579 $\pm$ 0.1000 & 0.9352 $\pm$ 0.0011 & 0.9354 $\pm$ 0.0011 \\
    CNN                             & 0.6569 $\pm$ 0.0958 & 0.9348 $\pm$ 0.2396 & 0.9829 $\pm$ 0.0035 & 0.9834 $\pm$ 0.0033 \\
    CNN-BLSTM                       & 0.4422 $\pm$ 0.0440 & 0.5087 $\pm$ 0.0994 & 0.9885 $\pm$ 0.0018 & 0.9887 $\pm$ 0.0017 \\
    \midrule
    \vspace{1pt}
    \textbf{QASE-net} & \textbf{0.4113 $\pm$ 0.0894} & \textbf{0.3731 $\pm$ 0.1492} & \textbf{0.9940 $\pm$ 0.0015} & \textbf{0.9941 $\pm$ 0.0014} \\
    \bottomrule
    \end{tabular}
\end{table*}

Fig.~\ref{fig:mse} illustrates the average MSE of all SNR prediction methods under different integer SNR inputs. QASE-net performs better across all SNRs than other models, confirming that CNN-BLSTM with an attention layer can provide more precise estimations for sEMG signals over different SNR levels. It is also noted that the MSE of the WL+MLP is comparable to other methods at SNRs below -12 dB. However, as the SNR increases, the MSE for WL+MLP also increases, unlike other methods that show different trends. As a result, WL+MLP exhibits relatively high MSE in terms of overall performance.

In Fig.~\ref{fig:scatter}, the upper and lower panels show the scatter plots of SNR predictions on the testing set by the WL+MLP and QASE-net, respectively. By comparing the two panels, the estimations from QASE-net are more tightly clustered around the actual SNR values than WL+MLP, especially at higher SNR levels (SNR above -12 dB). To better understand the statistical performance between two SNR prediction models, Fig.~\ref{fig:box} further employs box plots that provide a more precise visualization of the mean and quartiles of predictions under each integer SNR level. By comparing the interquartile ranges of both methods, the predictions from QASE-net are more concentrated, affirming that QASE-net consistently provides more accurate and stable SNR predictions over WL+MLP.


The results above reveal a noticeable decline in the performance of the WL+MLP, particularly in high SNRs. We attribute such a decline to using real sEMG recordings in this study, as the noisy sEMG signals with a higher SNR contain a higher proportion of sEMG. Real-world sEMG recordings may introduce complexity and variability due to diverse physiological conditions among subjects and various types of movements. Such complexity was not considered in the simulated sEMG data of the previous study, which are band passed white Gaussian noise segments \cite{oo2020signal}.

\section{Conclusion}

In this study, we proposed a non-intrusive signal quality assessment for sEMG with ECG contamination, leveraging CNN-BLSTM with an attention mechanism to predict the SNR of noisy sEMG. Moreover, the proposed model could assess the sEMG quality from the raw waveform, eschewing the need for hand-crafted features used in the previous work, such as WL+MLP. This data-driven feature extraction technique enhances the method's robustness in real-world scenarios. The experimental results demonstrate that our developed SNR prediction exhibits high correlations ($>$0.99 LCC) and low errors ($<$0.40 MSE). Notably, the proposed method consistently outperforms others across a wide range of SNR inputs. Our future research intends to employ this model in developing robust SNR-aware ECG removal techniques. In addition, we plan to explore the integration of other distance-based metrics that can provide a comprehensive assessment of sEMG quality from diverse perspectives. The holistic approach to sEMG quality assessment holds promise for enhancing the reliability and effectiveness of sEMG-based applications.
\newpage

\bibliographystyle{IEEEbib}
{
\bibliography{refs}
}

\end{document}